\documentclass[12pt]{amsart}
\usepackage[utf8]{inputenc}
\usepackage[T1]{fontenc}
\usepackage[margin=1.2in]{geometry}
\usepackage{tikz}
\usetikzlibrary{positioning}
\usepackage{xurl}
\usepackage{xfrac}
\usepackage{mathrsfs}
\usepackage{dsfont}
\usepackage{amsmath}
\usepackage{amsthm}
\usepackage{caption}
\usepackage{subcaption}
\usepackage{graphicx}
\usepackage{amssymb}
\usepackage{hyperref}
\usepackage{booktabs}
\usepackage{mathtools, cases}

\usepackage{multicol}
\usepackage[bibstyle=ieee, citestyle=numeric, backend=bibtex, sortcites=false, sorting=nyt, maxnames=5]{biblatex}
\DeclareNameAlias{default}{family-given}

\usepackage{hyperref}
\usepackage{xcolor}

\bibliography{../../LateX/biblio}

\newtheorem{theorem}{Theorem}
\newtheorem{lemma}[theorem]{Lemma}
\newtheorem{proposition}{Proposition} 

\theoremstyle{definition}
\newtheorem*{assumption-definition}{Assumption/Definition}
\newtheorem{definition}{Definition}
\newtheorem{remark}{Remark}
\newtheorem{example}{Example}

\newcommand{\calB}{\mathcal{B}}
\newcommand{\calC}{\mathcal{C}}
\newcommand{\calE}{\mathcal{E}}
\newcommand{\calF}{\mathcal{F}}

\newcommand{\calN}{\mathcal{N}}

\newcommand{\bbR}{\mathbb{R}}

\newcommand{\domSv}{\hspace{4pt} \text{dom}_S^v \hspace{4pt}}
\newcommand{\domS}{\hspace{4pt} \text{dom}_S \hspace{4pt}}

\newcommand{\dom}{\hspace{4pt} \text{dom} \hspace{4pt}}

\AtBeginEnvironment{example}{%
  \pushQED{\qed}%
}
\AtEndEnvironment{example}{\popQED\endexample}

\begin{document}

\title{On the formation of steady coalitions}

\author{Dylan Laplace Mermoud}
\address{Unit{\'e} de Math{\'e}matiques Appliqu{\'e}es, ENSTA Paris, Institut Polytechnique de Paris, 91120 Palaiseau, France}
\email{dylan.laplace.mermoud@gmail.com}

\date{\today}

\subjclass[2020]{MSC Primary 91A12; Secondary 91B24}

\begin{abstract}
  This paper studies the formation of the grand coalition of a cooperative game by investigating its possible internal dynamics. Each coalition is capable of forcing all players to reconsider the current state of the game when it does not provide sufficient payoff. Different coalitions may ask for contradictory evolutions, leading to the impossibility of the grand coalition forming. In this paper, we give a characterization of the impossibility, for a given state, of finding a new state dominating the previous one such that each aggrieved coalition has a satisfactory payoff. To do so, we develop new polyhedral tools related to a new family of polyhedra, appearing in numerous situations in cooperative game theory. 
\end{abstract}

\maketitle

\section{Introduction}

John von Neumann's first motivation while writing \citetitle{von1944theory}~\cite{von1944theory}, together with Oskar Morgenstern, was ``to embrace the phenomena of social equilibrium and change'' and ``the idea of capturing social configurations'' (\textcite{leonard2010neumann}). To this end, they defined the \emph{stable sets}, at that time simply called \emph{solutions}, to model the standards of behavior dictated by the game under consideration. These solutions concepts are based on the notion of \emph{domination}, a binary relation on the set of possible outcomes of the game. When one outcome dominates another, a coalition can force all the players into agreeing on the dominating one. In general, a given outcome is dominated by several others, and several coalitions, possibly overlapping yet with different interests, can subsequently force changes in opposite directions. Hence, the emergence of cooperation is particularly uncertain, and the ultimate outcome of this bargaining is even less foreseeable.

\medskip 

The unreasonableness of \textit{a priori} assuming cooperation between all the players gives rise to an even more challenging mathematical analysis. In the book ``Open Problems in Mathematics'' surveying most of the major mathematical problems of the century including Millennium problems such as the Navier-Stokes equations, the Hodge conjecture, the Riemann hypothesis and the P\hspace{2pt}=\hspace{2pt}NP problem, Nobel Prize winner Eric Maskin~\cite{maskin2016can} discusses the relevance of cooperative game theory to economics. He argues that ``cooperative theory seems to offer the important advantage of giving insight into how the coalitions behave'' and that ``cooperative games are more robust than noncooperative games''. However, according to him, more emphasis should be put on the externalities, and, indeed, the assumption that the grand coalition always forms should be abandoned. He finishes his essay by stating ``In my view, it remains an open problem---perhaps the most important open problem in cooperative theory---to develop an approach that properly accommodates the formation of multiple coalitions''. It is exactly in this spirit that the work presented in this paper was done. 

\medskip 

In this work, the social configurations are modelled by \emph{balanced collections}. They were introduced independently by \textcite{bondareva1963some} and \textcite{shapley1967balanced} in their respective proof of the Bondareva-Shapley Theorem. Balanced collections are families of positively weighted coalitions such that the sum of the weights of the coalitions that contain a given player is one, and this must be satisfied for every player. These weights represent the time spent by the players in the corresponding coalition and each player is active for exactly one unit of time. Balanced collections are a fundamental tool of cooperative game theory, as they are used in many characterizations of solution concepts such as the core or the nucleolus. Moreover, they generalize the \emph{coalition structures} introduced by \textcite{aumann1974cooperative} as a final configuration of the interaction between the players of the game. Furthermore, the balanced collections and their closely related counterparts, the \emph{unbalanced collections}, are closely related to domination, especially when it is via several coalitions simultaneously, which is in line with the critics of \textcite{maskin2016can} and allows for a mathematical study of the dynamics of the social configurations formed by the players.

\medskip 

Dually, the social context imposed on the players is represented by a cooperative game with transferable utility, or simply \emph{game}. A game is a set function \(v\) that associates a real number to each set of players, and 0 to the empty set.  The number \(v(S)\), called the \emph{worth} of coalition \(S\), represents the cumulative amount of utility the players in \(S\) can acquire by cooperating, in the worst possible case. Therefore, the grand coalition \(N\) can only be formed if there exists a way of allocating \(v(N)\) to the players such that, for each coalition \(S\), the sum of the utility allocated to the players in \(S\) is at least as large as the worth \(v(S)\). If this is not the case, it is reasonable to assume that the players for which the cumulated payment is smaller than the worth of the coalition they can form prefer to act on their own, independently of the rest of the players. The set of allocations satisfying every coalition is called the \emph{core}, and its nonemptiness is understood as an only necessary condition for global cooperation. 

\medskip 

The social context modeled by the coalition function generates a different quantity of utility allocated to the whole society of players according to how they organize themselves. It is possible to evaluate it by extending the coalition function \(v\) to the set of balanced collections. We do so by defining the worth of a balanced collection \( \calB \) as \( v(\calB) = \sum_{S \in \calB} \lambda_S v(S) \) with \( \lambda_S \) being the weight of \( S \in \calB \). Indeed, in the social configuration described by the balanced collection \( \calB \), the coalition \(S\) is active for a duration of \( \lambda_S \), hence generating an amount of utility \(\lambda_S v(S)\). Summing over all coalitions \( S \in \calB \) then yields the total utility of the balanced collection.\ \textcite{bondareva1963some} and \textcite{shapley1967balanced} have proven that the core of a game is nonempty when the grand coalition is one of the most efficient social configurations. Accordingly, to allow for global cooperation, we need to check whether the worth of the grand coalition is greater than the total utility of any balanced collection, for which the author, Grabisch and Sudh{\"o}lter~\cite{laplace2023minimal} developed an algorithm.

\medskip 

The present paper aims to provide several tools for a mathematical analysis of the stability of social organizations by taking into account a social context described by a game and an initial allocation. More specifically, we address the question of the formation of the grand coalition, provided that the core is nonempty. To examine the possibility of cooperation, we study the possible improvement, through domination, of the payment of several coalitions simultaneously by choosing a new state which can be enforced. We thus not only investigate whether global cooperation \emph{can} occur, but also \emph{how} we can achieve it.

\medskip 

In Section~\ref{section: preliminaries} we give the rigorous definitions of the concepts mentioned earlier. Section~\ref{section: characterization-stability} is devoted to the exposition of the particular angle this paper is adopting regarding the formation of the grand coalition. In Section~\ref{section: cooperahedra}, we present a new family of polyhedra, called \emph{cooperahedra}, related to the permutahedra and removahedra, which is regularly used in the proofs of the results from the next sections. Section~\ref{section: blind-spots} is devoted to the study of initial states that allow, or not, the emergence of global cooperation. We do so by giving a complete characterization of the states that are dominated by some others that give satisfactory payments to the discontent coalitions. Section~\ref{section: market} finishes this study by applying these results to market games.  

\section{Preliminaries}\label{section: preliminaries}

In this paper, we focus on a finite set of \emph{players} of cardinality \(n\), denoted by \(N\), and we want to know whether the players in \(N\) are willing to fully cooperate. For that to be the case, forming the coalition \(N\) must be in the interest of every player. In particular, there should not be any subcoalition of \(N\) that could provide to its players more than what \(N\) can provide to them. To formally quantify this idea, we use a \emph{coalitional function}, a set function \(v\) from \(2^N = \{S \mid S \subseteq N\}\) to \( \bbR \), which associate to each subset \(S\) of \(N\) a number \(v(S)\), called the \emph{worth} of \(S\). We interpret the worth of \(S\) as the amount of utility that the players in \(S\) have by cooperating during one unit of time, and that they can freely allocate among themselves. The set of nonempty subsets of \(N\) is written \(\mathcal{N}\) and its elements are called the \emph{coalitions}. 

\begin{definition}[\textcite{von1944theory}] \leavevmode \newline
  A game is an ordered pair \((N, v)\) where 
  \begin{enumerate}
    \item[\( \circ \)] \(N\) is a finite set of \emph{players}, called the \emph{grand coalition},
    \item[\( \circ \)] \(v\) is a set function \( v: 2^N \to \bbR \) such that \(v(\emptyset) = 0\). 
  \end{enumerate}
\end{definition}

As previously discussed, a necessary condition for the grand coalition to form is that it must be able to distribute more utility among the players than any of its subcoalitions can. Denote by \(X(v)\) the affine hyperplane of \(\bbR^N\) defined by 
\[
X(v) = \{x \in \bbR^N \mid x(N) = v(N)\}, 
\]
where \(x(S) = \sum_{i \in S} x_i\) for any coalition \(S \in \mathcal{N}\) and \(\bbR^N\) denotes the \(n\)-fold Cartesian product of \(n\) copies of \( \bbR \), one for each player. For any player \(i \in N\), the coordinate \(x_i\) represents the \emph{payment} of \(i\) by \(x\). The payment of a coalition \(S\) is simply the sum \(x(S)\) of the payment of the players. We denote by \(\mathbf{1}^i\) the \(i\)-th vector of the canonical base of \(\bbR^N\), i.e., the vector giving a payment of \(1\) to player \(i\) and \(0\) to the other players. We denote by \(\mathbf{1}^S\) the sum of vectors \(\mathbf{1}^S \coloneqq \sum_{i \in S} \mathbf{1}^i\). 

\medskip 

The set \(X(v)\) represents all the possible ways to allocate \(v(N)\) among \(n\) players. The elements of \(X(v)\) are called the \emph{preimputations} of the game \((N, v)\), and they represent the possible \emph{states} of a given game. 

\medskip 

Therefore, the question is whether there exist some preimputations \(x\) such that the payment of any coalition \(S\) by \(x\) is better than what \(S\) can achieve on its own. In other words, we want to know whether the following set 
\[
C(v) = \{x \in X(v) \mid x(S) \geq v(S), \forall \hspace{1pt} S \in \calN \}~,
\]
called the \emph{core} of the game \((N, v)\), is nonempty. This question was solved by \textcite{bondareva1963some} and \textcite{shapley1967balanced} independently, using the concept of \emph{balanced collections}. 

\begin{definition}[\textcite{bondareva1963some}, \textcite{shapley1967balanced}] \leavevmode \newline 
  A collection of coalitions \( \calB \subseteq \calN \) is \emph{balanced} on \(N\) if there exists a set of positive weights \( \lambda = \{\lambda_S \mid S \in \calB \} \) such that \(\sum_{S \in \calB} \lambda_S \mathbf{1}^S = \mathbf{1}^N\). 
  \end{definition}
  
A balanced collection \( \calB \) models a social configuration of the players in \(N\). They represent how the players can be organized, respecting two natural conditions:
\begin{enumerate}
  \item[\( \circ \)] The players forming a coalition spend the same time in it, 
  \item[\( \circ \)] Each player is `active' for exactly one unit of time.
\end{enumerate}
The weights \( \lambda_S \) represent the fraction of the time that the players spend in the coalition \(S\). Indeed, we can reformulate the equality in the definition above by 
\[
\sum_{\substack{S \in \calB \\ S \ni i}} \lambda_S = 1, \qquad \forall \hspace{1pt} i \in N.
\]
The total worth of the balanced collection \( \calB \) is determined by \( \sum_{S \in \calB} \lambda_S v(S) \). Indeed, each coalition \(S\) has a worth of \(v(S)\) and is active for \(\lambda_S\) units of time.

\begin{example}\label{ex: bal-coll}
  Consider the game defined on \( N = \{a, b, c\} \) by \( v(S) = 0 \) if \( \lvert S \rvert \in \{0, 1\} \), \(v(S) = 0.8\) if \(\lvert S \rvert = 2\) and \(v(N) = 1\). We observe that the grand coalition is the one with the largest value and that any partition of it has a worth not exceeding \(v(N)\). But the players can still be organized in a way that gives them more utility than \(v(N)\). Indeed, if each player spends half of its time with another player, and the second half of its time with the last player, during each half, the coalition secures a utility of \(0.4\), and this three times. In this case, the balanced collection is 
  \[
  \calB = \{ \{a, b\}, \{a, c\}, \{b, c\} \} \text{ with weights being } \lambda_{\{a, b\}} = \lambda_{\{a, c\}} = \lambda_{\{b, c\}} = \frac{1}{2}
  \]
  and \(\sum_{S \in \calB} \lambda_S v(S) = \frac{3}{2} \cdot 0.8 = 1.2 > 1\). 
\end{example}

The simple, yet rich, definition of the balanced collections made them of specific interest in other fields of mathematics, mostly in combinatorics, under the name of regular hypergraphs, or perfect fractional matching. They are also closely related to uniform hypergraphs, both the resonance and the braid hyperplane arrangements \cite{EPTCS403.27}. 

\medskip 

In Example~\ref{ex: bal-coll}, we cannot expect the grand coalition \(N\) to form, even if it has the largest worth and no partition of it exceeds its worth. The players have a better way to be organized than forming a non-decomposable block \(N\). 

\medskip 

A game is \emph{balanced} if the balanced collection \( \{N\} \) is one with the maximal worth. In other words, a game \( (N, v) \) is balanced if, for all balanced collection \( \calB \), we have 
\[
\sum_{S \in \calB} \lambda_S v(S) \leq v(N). 
\]

\begin{theorem}[\textcite{bondareva1963some}, \textcite{shapley1967balanced}]\label{th: bondareva-shapley} \leavevmode \newline
The core of a game \( (N,v) \) is nonempty if and only if \( (N, v) \) is balanced. 
\end{theorem}

To put it differently, the existence of an allocation of \(v(N)\) among the players such that all coalitions receive a satisfactory payment, is equivalent to \( \{N\} \) being one of the most efficient organizations for the players in \(N\). 

\medskip 

However, even if solutions that benefit everyone exist, they may not be accessible from the current state of the game. Assume that the players' initial endowments are the payments of an initial state represented by the preimputation \(x\). Several subcoalitions of \(N\) can be unsatisfied by this payment but may not be able to coordinate their responses, because their interests are too divergent. Considering that the sum of all the payments must not change, an increase in the payments of some players induces a decrease in the payments of other players, leading to the inability for these coalitions to agree on a counterproposal for \(x\) and hence improve their situations simultaneously. At this stage, nothing ensures that the grand coalition \(N\) will form. 

\medskip 

Naturally, according to the payment they receive, the coalitions may prefer a preimputation \(x\) to a preimputation \(y\) if, for example, each player in the coalition gets a strictly better payment at \(x\) than at \(y\). On the other hand, a preimputation \(x\) is affordable for a coalition \(S\) if, by working on their own, they can produce the \( \lvert S \rvert \)-dimensional vector \( x_{|S} = {(x_i)}_{i \in S} \), i.e., if \(x(S) \leq v(S)\). 

\begin{definition}[\textcite{von1944theory}]\label{def: domination} \leavevmode \newline
  Let \((N, v)\) be a game, \(x, y\) be two preimputations of \((N, v)\) and let \( S \in \calN \) be a coalition. We say that \(x\) dominates \(y\) via \(S\) with respect to \((N, v)\), denoted \(x \domSv y\) if
  \begin{enumerate}
    \item\label{item: affordable} \(x\) is \emph{affordable} to \(S\) w.r.t. \((N, v)\), i.e, \(x(S) \leq v(S)\), 
    \item \(x\) \emph{improves} \(y\) on \(S\), i.e., \(x_i > y_i\) for all \(i \in S\). 
\end{enumerate}
\end{definition}
  
If the considered game \((N, v)\) is clear from the context, we simply write \(x \domS y\). We say that \(x\) \emph{dominates} \(y\), denoted \(x \dom y\), if there exists \( S \in \calN \) such that \(x \domS y\). 

\section{How to characterize the steadiness of a coalition?}\label{section: characterization-stability}

The domination relation induces a dynamic on the set of preimputations, where the initial state of the game evolves towards dominating preimputations. 

\medskip 

Indeed, if a preimputation \(x\) is dominated by \(y\) via the coalition \(S\), we must have that \(x(S) < y(S) \leq v(S)\), hence the players in \(S\) are not satisfied by their payment, and therefore want to renegotiate the state of the game. 

\medskip 

While bargaining with the whole set of players, the coalition \(S\) expresses its dissatisfaction about the current state \(x\), and demands that they all collectively choose the preimputation \(y\) as the new state. Their main argument is that they can obtain a payment of \(v(S) > x(S)\) by themselves, hence they threaten to defect and leave the grand coalition \(N\). 

\medskip 

Even if the players in \(N \setminus S\) necessarily have a better payment at \(x\) rather than at \(y\), they still prefer to agree on \(y\) than seeing the coalition \(S\) leaving the grand coalition. 

\begin{theorem}\label{th: force-pay}
  Let \((N, v)\) be a balanced game, let \(x, y\) be two preimputations and let \(S\) be a coalition such that \(y \domS x\). The coalition \(N \setminus S\) receives a payment at \(y\) larger than the worth of any balanced collection on \(N \setminus S\).
\end{theorem}

\begin{proof}
  First, notice that each balanced collection on \(N \setminus S\) is a balanced collection on \(N\) when coalition \(S\) is added, and each balanced collection on \(N\) containing the coalition \(S\) with a weight of \(1\) becomes a balanced collection on \(N \setminus S\) when the coalition \(S\) is put aside. By balancedness of \((N, v)\), for each balanced collection \( \calB \) on \(N\) containing coalition \(S\) with a weight \(\lambda_S = 1\), we have \( v(S) + \sum_{T \in \calB \setminus \{S\}} \lambda_T v(T) \leq v(N) \), or equivalently, \(\sum_{T \in \calB'} \lambda_T v(T) \leq v(N) - v(S)\), with \(\calB'\) being any possible balanced collection on \(N \setminus S\). Then, if the players in \(N \setminus S\) have to work without the help of players in \(S\), they would not be able to get more than \(v(N) - v(S)\). However, because \(x\) is affordable to \(S\) (Item~\ref{item: affordable} of Definition~\ref{def: domination}), the payment of \(S\) at \(x\) cannot exceed \(v(S)\), and the payment of \(N \setminus S\) at \(x\) satisfies \( x(N \setminus S) = x(N) - x(S) = v(N) - x(S) \geq v(N) - v(S) \), and is indeed greater than what they could obtain on their own. 
\end{proof}

Hence, even if the payment of \(N \setminus S\) decreases by changing the current state from \(x\) to \(y\), they have no other choice than to accept this new state \(y\). Indeed, the players in \(N \setminus S\) can be organized following the configuration described by any balanced collection on \(N \setminus S\), but the worth of these never exceed \(y(N \setminus S)\). 

\medskip 

Thus, both \(S\) and \(N \setminus S\) will accept \(y\). Nevertheless, there can still exist a third coalition \(T\) which is unsatisfied with this new preimputation, so these players are not yet likely to accept \(y\) as a final allocation of the whole worth \(v(N)\). 

\medskip 

Following this idea, an agreement on a final state can only be achieving by choosing a state belonging to the core.

\begin{definition}[\textcite{von1944theory}] \leavevmode \newline 
  A set \(K\) of \(X(v)\) is a \emph{stable set} if it is 
  \begin{enumerate}
    \item[\( \circ \)] \emph{internally stable}, i.e., for all \(z \in K\), there exists no \(x \in K\) such that \(x \dom z\),
    \item[\( \circ \)] \emph{externally stable}, i.e., for all \(y \in X(v) \setminus K\), there exists \(x \in K\) such that \(x \dom y\).
  \end{enumerate}
\end{definition}

We say that the core of a game is \emph{stable} if it is a stable set, or equivalently if it is externally stable. Indeed, the preimputations in the interior of the core cannot be affordable for any coalition, and the preimputation on the frontier of the core cannot improve the payment of all the players forming the coalition for which it is affordable. 

\medskip 

The stable sets were the first solution concept for cooperative games. Despite their appealing definition, using the stable sets is difficult, as there may be none, several, or even a continuum of them. Moreover, they are not convex, and therefore cannot be easily described~\cite{lucas1969proof, lucas1992neumann}. However, if the core is externally stable, it is the unique stable set~\cite{driessen2013cooperative}, and therefore the properties of both the core and the stable sets are satisfied by a unique polytope of preimputations. 

\medskip 

By definition, when the core is (externally) stable, any preimputation \(x\) outside the core is dominated by an element inside the core, say \(z\). Therefore, a coalition \(S\) which is discontent with its payment at \(x\), can demand to change the current state to \(z\). Thanks to Theorem~\ref{th: force-pay}, the players not in \(S\) cannot prevent this for happening. Since the state \(z\) is in the core, there exists no coalition unsatisfied with the new state and no one is threatening to defect from \(N\). Hence, the grand coalition is likely to form. 

\begin{definition}
  Let \( (N, v) \) be a balanced game and \(x\) be the initial state of the game. We say that \(N\) will form if \(x\) belongs to the core, or if \(x\) is dominated by a core element.
\end{definition}

According to this definition, the grand coalition of the game \( (N, v) \) forms for any initial preimputation if and only if the core is stable. 

\medskip 

To further illustrate the concept of domination, we consider \emph{flow games}, defined by \textcite{kalai1982totally}. Let \(G = (V, E)\) be a directed graph, with \(V\) being the set of vertices containing a \emph{source node} \(s\) and a \emph{target node} \(t\), and \(E\) being the set of (directed) edges. Alongside this graph, we define two functions: 
\begin{enumerate}
  \item[\( \circ \)] \(c: E \to \bbR_+\) associates each edge with its \emph{capacity}, 
  \item[\( \circ \)] \(p:E \to N\) associates each edge with the player owning it.
\end{enumerate}

\medskip 

\noindent A \emph{flow} \(f\) through a graph \(G\) is a mapping \(f: E \to \bbR_+\) subject to two constraints:
\begin{enumerate}
  \item[\( \circ \)] For every edge \(e \in E\), we have 
  \[
  f(e) \leq c(e),
  \] 
  \item[\( \circ \)] For each vertex \(v \in V\) apart from \(s\) and \(t\), the following equality holds
  \[
  \sum_{e \in E_v} f(e) = \sum_{e' \in E^v} f(e'),
  \]   
  where \( E_v \) (\(E^v\)) is the set of edges having \(v\) as source (target) node.
\end{enumerate}
The second condition in this definition simply states that the amount of flow entering a node is equal to the amount of flow leaving it. The value of a flow is defined by 
\[
\lvert f \rvert = \sum_{e \in E_s} f(e) = \sum_{e' \in E^t} f(e'). 
\]
For a coalition \( S \in \calN \), let \(G_S\) be the subgraph of \(G\) restricted to the edges owned by the players in \(S\). The coalition function \(v\) of the flow game \((N, v)\) associated with \(G\) maps \(S\) to the maximal amount of flow carried throughout \(G_S\) between \(s\) and \(t\). 

\medskip 

Flow games are convenient examples for our discussion because they are presented in a more compact way than the exhaustive enumeration of the worth of an exponential number of coalitions, and their cores are always nonempty \cite{kalai1982totally}.

\begin{figure}[ht]
  \begin{center}
    \begin{tikzpicture}[node distance={23mm}, thick, main/.style = {draw, circle}] 
      \node[main] (s) {$s$}; 
      \node[main] (1) [right of=s] {$1$}; 
      \node[main] (2) [below right of=s] {$2$}; 
      \node[main] (3) [right of=1] {$3$};
      \node[main] (4) [below right of=1] {$4$};
      \node[main] (5) [below right of=3] {$5$};
      \node[main] (t) [right of=3] {$t$};
      
      \draw[->] (s) to node[midway, above] {\small $6, a$} (1);
      \draw[->] (s) to node[midway, below left] {\small $6, c$} (2);
      \draw[->] (1) to node[midway, above] {\small $6, d$} (3);
      \draw[->] (1) to node[midway, right] {\small $\; 4, a$} (2);
      \draw[->] (2) to node[midway, below] {\small $4, b$} (4); 
      \draw[->] (3) to node[midway, above] {\small $4, b$} (t);
      \draw[->] (4) to node[midway, left] {\small $8, c \;$} (3);
      \draw[->] (4) to node[midway, below] {\small $6, a$} (5);
      \draw[->] (3) to node[midway, right] {\small $\hspace{1pt}4, d$} (5);
      \draw[->] (5) to node[midway, right] {\small $\; 6, a$} (t);
    \end{tikzpicture} 
  \end{center}
  \caption{Directed graph describing a flow game on \( N = \{a, b, c, d\} \).}\label{fig: flow}
\end{figure}
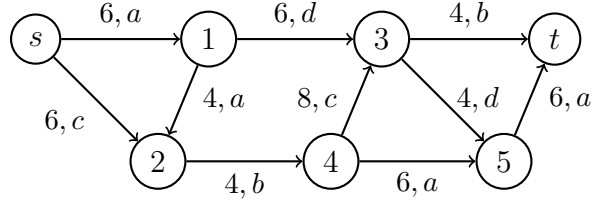

Following the example depicted in Figure~\ref{fig: flow}, the coalition function of the associated game flow is defined by 

\medskip 

\begin{center}
\begin{tabular}{rcccccccc}
\(S\): \quad & \( \{a, b\} \) & \( \{a, d\} \) & \( \{b, c\} \) & \( \{a, b, c\} \) & \( \{a, b, d\} \) & \( \{a, c, d\} \) & \( \{b, c, d\} \) & \(N\) \\
\midrule
\( v(S) \): \quad & \(4\) & \(4\) & \(4\) & \(4\) & \(6\) & \(4\) & \(4\) & \(10\)
\end{tabular}
\end{center}

\medskip 

\noindent and \(v(S) = 0\) otherwise. The flow corresponding to the grand coalition \(N\) is 

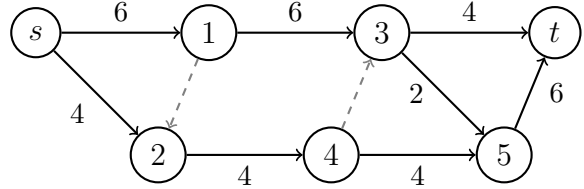
\begin{center}
  \begin{tikzpicture}[node distance={23mm}, thick, main/.style = {draw, circle}, row sep = 17mm] 
    \node[main] (s) {$s$}; 
    \node[main] (1) [right of=s] {$1$}; 
    \node[main] (2) [below right of=s] {$2$}; 
    \node[main] (3) [right of=1] {$3$};
    \node[main] (4) [below right of=1] {$4$};
    \node[main] (5) [below right of=3] {$5$};
    \node[main] (t) [right of=3] {$t$};
    
    \draw[->] (s) to node[midway, above] {\small $6$} (1);
    \draw[->] (s) to node[midway, below left] {\small $4$} (2);
    \draw[->] (1) to node[midway, above] {\small $6$} (3);
    \draw[->, dashed, gray] (1) to (2);
    \draw[->] (2) to node[midway, below] {\small $4$} (4); 
    \draw[->] (3) to node[midway, above] {\small $4$} (t);
    \draw[->, dashed, gray] (4) to (3);
    \draw[->] (4) to node[midway, below] {\small $4$} (5);
    \draw[->] (3) to node[midway, left] {\small $2\hspace{3pt}$} (5);
    \draw[->] (5) to node[midway, right] {\small $\; 6$} (t);
  \end{tikzpicture} 
\end{center}
with the dashed edges carrying no flow. The number associated with each edge is the flow transiting through it and respects its capacity constraints. To carry as many as \(10\) units of flow, all the players had to collaborate and form \(N\).

\medskip 

Let us assume that since the last agreement between the players, the capacities of each edge may have changed, but that the maximal flow is still \(10\), and that the current payments of the players are given by the preimputation \(x = (1, 4, 4, 1)\). 

\medskip 

We notice that this preimputation no longer belongs to the core, because coalition \( \{a, d\} \) has a worth of \(4\) and only receives a payment of \(2\). Therefore, the coalition \( \{a, d\} \) demands that its payment is reevaluated, or it will shut down all the edges belonging to its players. If coalition \( \{b, c\} \) relies only on its edges, it can only carry a flow of \(4\). 

\medskip 

By agreeing with coalition \( \{a, d\} \) on a preimputation dominating \(x\) via \( \{a, d\} \) which is no longer dominated via the same coalition, say \(y = (2, 3, 3, 2)\), it secures a payment of \( y(\{b, c\}) = 6 \), which is larger than what they can achieve on their own. Moreover, \(y\) belongs to the core, so the preimputation \(y\) is satisfactory for every coalition. 

\medskip 

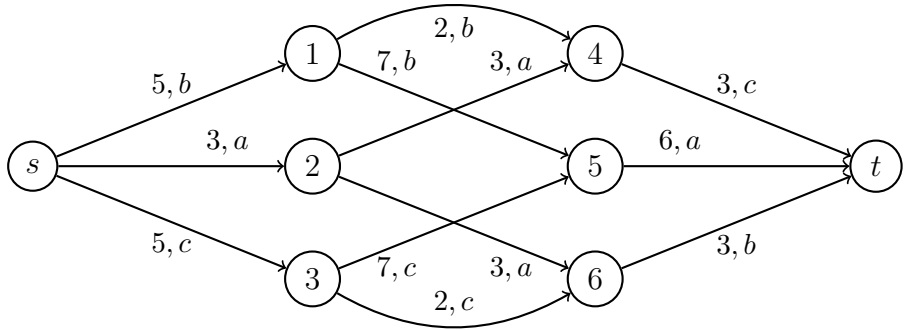
\begin{figure}[ht]
  \begin{center}
    \begin{tikzpicture}[node distance={15mm}, thick, main/.style = {draw, circle}, row sep = 12mm] 
    \node[main] (s) {$s$}; 
    \node[main] (2) [right = 3cm of s] {$2$}; 
    \node[main] (1) [above of=2] {$1$}; 
    \node[main] (3) [below of=2] {$3$};
    \node[main] (5) [right = 3cm of 2] {$5$};
    \node[main] (4) [above of=5] {$4$};
    \node[main] (6) [below of=5] {$6$};
    \node[main] (t) [right = 3cm of 5] {$t$};

    \draw[->] (s) to node[midway, above] {\small $5, b$} (1);
    \draw[->] (s) to node[near end, above] {\small $3, a$} (2);
    \draw[->] (s) to node[midway, below] {\small $5, c$} (3);

    \draw[->] (1) to[out=30, in=150] node[midway, below] {\small $2, b$} (4);
    \draw[->] (1) to node[near start, above] {\small $7, b$} (5);
    \draw[->] (2) to node[near end, above] {\small $3, a$} (4);
    \draw[->] (2) to node[near end, below] {\small $3, a$} (6);
    \draw[->] (3) to node[near start, below] {\small $7, c$} (5);
    \draw[->] (3) to[out=-30, in=-150] node[midway, above] {\small $2, c$} (6);

    \draw[->] (4) to node[midway, above] {\small $3, c$} (t);
    \draw[->] (5) to node[near start, above] {\small $6, a$} (t);
    \draw[->] (6) to node[midway, below] {\small $3, b$} (t);
    \end{tikzpicture} 
  \end{center}
\caption{Directed graph describing a flow game on \( N = \{a, b, c\} \).}\label{fig: flow-2}
\end{figure}

However, it is not always possible to have a core element dominating the current preimputation, even when the core is nonempty. Let us consider the directed graph described in Figure~\ref{fig: flow-2}. 

In this example, the coalition function of the associated game flow is defined by 

\medskip 

\begin{center}
\begin{tabular}{rccccccc}
\(S\): \quad & \( \{a\} \) & \( \{b\} \) & \( \{c\} \) & \( \{a, b\} \) & \( \{a, c\} \) & \( \{b, c\} \) & \(N\) \\
\midrule
\(v(S)\): \quad & \(0\) & \(0\) & \(0\) & \(8\) & \(8\) & \(4\) & \(12\)
\end{tabular}
\end{center}

\medskip 

\noindent Assume that the current state is \(x = (2, 5, 5)\). Coalition \( \{a, b\} \) is not satisfied with its payment, therefore \(x\) does not belong to the core. The coalition \( \{a, b\} \) threatens to shut down its edges, forcing player \(c\) to renegotiate the preimputation. By Theorem~\ref{th: force-pay}, player \(c\) knows that he has to give up some utility and access the demand of coalition \( \{a, b\} \), say on preimputation \(y = (2.5, \hspace{1pt} 5.5, \hspace{1pt} 4)\).

\medskip 

However, coalition \( \{a, c\} \) is not content with its new payment \( y(\{a, c\}) = 6.5 \), which is even worse than \( x(\{a, b\}) = 7 \) that was already not satisfactory. In fact, any preimputation dominating via \( \{a, b\} \) decreases the payment of \( \{a, c\} \), and vice versa. For this game, a renegotiation of the state \(x\) supported by domination that leads to a new state which is satisfactory for every coalition is impossible, even though the core is nonempty (it contains, for instance, the preimputation \( (6, 3, 3) \)). 

\medskip 

To summarize, the main assumption we use in this paper is that a set of players \(N\) agree to cooperate if the core of the game \( (N, v) \), which describes the power relations between all the subcoalitions of \(N\), contains a state which dominates the initial state. Balancedness of this game is thus a necessary condition, but is not sufficient because the aggrieved coalitions may have divergent interests, and cannot agree on a coordinated counterproposal to the current state. In the next sections, we seek a characterization of simultaneous improvements of the payments of discontent coalitions by a dominating core element, but first, we introduce a new tool, created by mixing polyhedral theory with the Bondareva-Shapley Theorem.

\section{Cooperahedra}\label{section: cooperahedra}

The Bondareva-Shapley Theorem is a powerful and well-known algorithmic tool to check the nonemptiness of the core of a given game. On one hand, we can interpret this result as we did previously, namely by characterizing the equivalence for the grand coalition \(N\) between having the possibility to give a satisfactory payment to each of its subsets with its worth \(v(N)\) and being one of the most efficient social configuration. 

\medskip 

On the other hand, the Bondareva-Shapley Theorem builds a bridge between geometry and algorithmics. Indeed, it characterizes the nonemptiness of a given polyhedron by using a very simple algorithmic routine, consisting on the verification of a finite number of boolean values, one for each \emph{minimal} balanced collection. 

\medskip 

The author, Grabisch and Sudh{\"o}lter~\cite{laplace2023minimal} have converted this mathematical result into a working computer program checking the balancedness of a game given as an input. It outperforms classical linear programming algorithms, as long as we have access to the set of minimal balanced collections on the set of players we consider. 

\medskip 

A minimal balanced collection is a balanced collection such that no proper subcollection is balanced, and every balanced collection is the union of the minimal balanced collections it contains. Therefore, the worth \(\sum_{S \in \calB} \lambda_S v(S)\) of a balanced collection \( \calB \) is a convex combination of the worth of the minimal balanced collections of which it is the union, and then the Bondareva-Shapley Theorem still holds when we replace the balanced collections by the minimal balanced collections, which are much less numerous and much easier to generate. Using an algorithm based on Peleg's inductive method~\cite{peleg1965inductive}, the minimal balanced collections for any game have been generated for up to \(7\) players by the author, Grabisch and Sudh{\"o}lter~\cite{laplace2023minimal}.

\medskip 

Using the minimal balanced collections, it is possible to investigate additional geometrical properties of the core of a game. These geometrical properties depend on the set of the most efficient minimal balanced collections of the game, or, equivalently, they depend on the uniqueness of \( \{ N \} \) as a most profitable minimal balanced collection. 

\medskip 

Let \((N, v)\) be a balanced game. If \( \{N\} \) is the unique most efficient organization for the players in \(N\), we say that \(N\) is \emph{vital}. If there exists a balanced collection \( \calB \neq \{N\} \) which is as efficient as \( \{N\} \), i.e., for which there exists weights \( \{ \lambda_S \mid S\in \calB \} \) such that
\begin{equation} \label{eq: effective}
\sum_{S \in \calB} \lambda_S v(S) = v(N),
\end{equation}
the coalitions in \( \calB \) are called \emph{effective} coalitions w.r.t. \((N, v)\), and we denote their set by \( \calE(v) \). As a union of balanced collections, \( \calE(v) \) is itself a balanced collection. These coalitions determine the dimension of the core.

\begin{proposition}\label{prop: effective}
  Let \((N, v)\) be a balanced game. \(S\) is an effective coalition if and only if each core element \(x\) satisfies \(x(S) = v(S)\). 
\end{proposition}

\begin{proof}
Let \( \calB \) be a balanced collection with a system of weights \( \lambda = \{\lambda_S \mid S \in \calB \} \) satisfying Equation~\eqref{eq: effective}. Let \(x \in C(v)\). Then, 
\[
v(N) = x(N) = \sum_{S \in \calB} \lambda_S v(S) \geq \sum_{S \in \calB} \lambda_S v(S) = v(N). 
\]
Because for all \( S \in \calB \), we have \( \lambda_S > 0 \), hence \( x(S) = v(S) \).

\medskip 

For the reverse implication, let \( S \neq N \) be a coalition such that, for all core elements \(x\), we have \( x(S) = v(S) \). By contradiction, assume that \(S\) does not belong to a balanced collection \( \calB \) satisfying Equation~\eqref{eq: effective}. Then, by the Bondareva-Shapley Theorem, there exists \( \varepsilon > 0 \) such that the game \( (N, v^\varepsilon) \) that differs from \((N, v)\) only inasmuch as \( v^\varepsilon(S) = v(S) + \varepsilon \) is still balanced. Hence, for \(x \in C(v^\varepsilon)\), it follows that \( x(S) > v(S) \) and \( x \in C(v) \), then the desired contradiction has been obtained. 
\end{proof}  

The affine hull of the core is therefore 
\[
\text{aff}\big( C(v) \big) = \bigcap_{S \in \calE(v)} A_S(v) \qquad \text{with} \qquad A_S(v) \coloneqq \{x \in X(v) \mid x(S) = v(S)\}.
\]

We now have a complete characterization of the core nonemptiness and its dimension. The goal of this section is to extend this result to a new class of polyhedra. 

\medskip 

Let denote by \( \langle x, y \rangle \) the usual scalar product between \( x, y \in \bbR^N \). 

\begin{definition}\label{def: cooperahedron}
A \emph{cooperahedron} is a polyhedron \(P \subseteq \bbR^N\) of the form 
\[
P = \{ x \in \bbR^N \mid \langle e^N, x \rangle = b^0 \text{ and } \langle z^i, x \rangle \diamond_i b^i, \hspace{3pt} \forall i \in I_P \},
\]
where \(I_P\) is finite, and \( \forall i \in I_P\), we have \( \diamond_i \in \{ \geq, >, <, \leq \} \) and \( z^i \in {\{0, 1\}}^N \). 
\end{definition}

For convenience, we denote by \(I^\prec_P\) and \(I^\succ_P\) the sets of indices defined by 
\[
I^\prec_P = \{ i \in I_P \mid \diamond_i \in \{ <, \leq \} \} \quad \text{and} \quad I^\succ_P = \{ i \in I_P \mid \diamond_i \in \{ \geq, > \} \}. 
\]
If \(i \in I^\prec_P\), we write \(\prec_i\) instead of \(\diamond_i\), and if \(i \in I^\succ_P\), we write \(\succ_i\) instead of \(\diamond_i\). For each \(i \in I_P\), we can associate with \(z^i\) a coalition \(S_i\) defined by 
\[
  S_i = \begin{cases} 
    \{j \in N \mid z^i_j = 1\}, & \quad \text{if } i \in I^\succ_P, \\
    \{j \in N \mid z^i_j = 0\}, & \quad \text{if } i \in I^\prec_P.
  \end{cases}
\]
Let \(x \in \bbR^N\) be such that \(x(N) = \langle e^N, x\rangle = b_0\). For each \(i \in I^\prec_P\), the inequality \(\langle z^i, x \rangle \prec_i b^i\) becomes \(x(N \setminus S_i) \prec_i b^i\), which can be rewritten as \(-x(N \setminus S_i) \succ_i -b_i\), and because \(x(N) = b^0\), we get that \(\langle z^i, x \rangle \prec_i b^i \) is equivalent to \( x(S_i) \succ_i b^0 - b^i\).

\medskip 

Hence, for each cooperahedron, we can define a game \( (\calF_P, v_P) \) on a set system \( \calF_P \) which can be a proper subcollection of \( \calN_P \), by
\[
\calF_P = \{S_i \mid i \in I\} \qquad \text{and} \qquad v_P(S_i) = \begin{cases}
  b_i, & \quad \text{if } i \in I^\succ_P, \\
  b^0 - b^i, & \quad \text{if } i \in I^\prec_P.
\end{cases}
\]
To get the same description as in Definition~\ref{def: cooperahedron} we need to take the strict inequalities into account. Let \(I^*_P\) be the subset of \(I_P\) containing the indices \(i\) such that \( \diamond_i \in \{<,>\} \). To obtain the cooperahedron from the core of \((\calF_P, v_P)\), we need to remove the core elements which are affordable for the coalitions \( \{ S_i \mid i \in I^*_P \} \). This is where the characterization of nonempty cooperahedra stems from. 

\begin{theorem}\label{th: cooperahedron-non-empty}
  A cooperahedron is nonempty if and only if \((\calF_P, v_P)\) is balanced and 
  \[
  \calE(\calF_P, v_P) \cap \{S_i \mid i \in I^*_P\} = \emptyset.
  \] 
\end{theorem} 

\begin{proof} 
  Let \(P\) be a cooperahedron defined by
  \[
  P = \{x \in \bbR^N \mid \langle e^N, x \rangle = b^0 \text{ and } \langle z^i, x \rangle \diamond_i b^i, \hspace{3pt} \forall i \in I_P\}. 
  \]
  Assume that there exists \(x \in P\). Therefore, \(x\) belongs to the core of \((\calF_P, v_P)\), and is not affordable for the coalitions in \( \{ S_i \mid i \in I^*_P \} \). Thus, \((N, v)\) is balanced, and no coalition of \(I^*_P\) is effective, i.e., \( \calE(\calF_P, v_P) \cap \{ S_i \mid i \in I^*_P \} = \emptyset \).

  \medskip 

  Assume that \( (\calF_P, v_P) \) is balanced and that \( \calE(\calF_P, v_P) \cap \{S_i \mid i \in I^*_P\} = \emptyset \). Therefore, for each coalition in \( \{ S_i \mid i \in I^*_P \} \), there exists a core element \(x^i\) such that \( x^i(S_i) > v(S_i) \). Consequently, the midpoint \( \bar{x} = \frac{1}{\lvert I^*_P \rvert} \sum_{i \in I^*_P} x^i \) belongs to the core of \( (\calF_P, v_P) \) and for all \( i \in I^*_P\), we have \(\bar{x}(S_i) > v(S_i) \), hence belongs to \(P\). 
\end{proof}

\begin{example}
  Let \( x \in X(v) \). The set \( P = \{ y \in X(v) \mid y \domS x \} \) is a cooperahedron. Indeed, this cooperahedron can be rewritten as 
  \[
  P = \{ y \in X(v) \mid y(S) \leq v(S) \text{ and } y_i > x_i, \forall i \in S\}. 
  \]
  Notice the orientation on the constraint on \(S\). We therefore have that \( \calF_P = \{N \setminus S\} \cup S\), and, for all \( T \in \calF_P \), we have
  \[
  v_P(T) = \begin{cases}
    v(N) - v(S), & \quad \text{if } T = N \setminus S, \\
    x_i, & \quad \text{if } T \in S.
  \end{cases}
  \]
  Strict inequalities are required to describe the constraints involving the players in \(S\), then \( I^*_P = S\). The preimputation \(x\) is dominated via \(S\) if and only if \(P\) is nonempty, which by Theorem~\ref{th: cooperahedron-non-empty} is equivalent to 
  \[
  v_P(N \setminus S) + \sum_{i \in S} v_P(\{i\}) < v(N), 
  \]
  because \( \calF_P \) is itself a minimal balanced collection and intersects \(I^*_P\). This gives that \(x\) is dominated via \(S\) if and only if \(v(N) - v(S) + x(S) < v(N)\), i.e., \(v(S) > v(S)\). 
\end{example}

The class of cooperahedra includes several well-known classes of polytopes and polyhedra, for instance, the class of deformed permutohedra (\textcite{postnikov2009permutohedra}), which contains the quotientopes (\textcite{pilaud2019quotientopes}), the class of removahedra (\textcite{pilaud2017nestohedra}), which contains the permutreehedra (\textcite{pilaud2018permutrees}). All these polyhedra come from the standard permutohedron, which was already used in mathematical social science by \textcite{guilbaud1963analyse} who gave it its name. 

\medskip 

The \emph{(standard) permutohedron} is the convex hull of the \(n{!}\) points defined by permuting the coordinates of the vector \( (1, \ldots, n) \in \bbR^N \). We can also define it using a system of linear inequalities, as the core of the strictly convex game \((N, v)\) defined by 
\begin{equation} \label{eq: permutohedron-game}
v(S) = \binom{\lvert S \rvert + 1}{2} = \frac{\lvert S \rvert \left( \lvert S \rvert + 1\right)}{2}. 
\end{equation}
Many classes of polyhedra are defined from this fundamental polytope, for instance, the aforementioned deformed permutohedra and removahedra. A \emph{deformed permutohedron} is obtained from the permutohedron by parallel translation of its facets so that the directions of all edges are preserved. More formally, we have the following definition. Denote by \( \mathfrak{S}_N \) the group of permutations of \(N\). 

\begin{definition}[\textcite{postnikov2009permutohedra}] \leavevmode \newline 
  A \emph{deformed permutohedron} is the convex hull of \(n{!}\) points \(v_\pi \in \bbR^N\) labeled by \( \mathfrak{S}_N \) such that, for any \(\pi \in \mathfrak{S}_N\) and any adjacent transposition \(\tau_i = (i, i+1)\), we have 
  \[
  v_\pi - v_{\tau_i(\pi)} = \alpha_{\pi, i} \left( e^{\pi(i)} - e^{\pi(i+1)} \right), \qquad \text{where} \qquad \alpha_{\pi, i} \geq 0.
  \]
\end{definition}

The fact that a deformed permutohedron is a cooperahedron is implied by the Submodularity Theorem, which is a corollary of Rado's inequality~\cite{rado1952inequality}. 

\begin{theorem}[Submodularity Theorem] \leavevmode \newline
The deformed permutohedra of dimension at most \(n-1\) are in bijection with the supermodular set function on \(2^N\) vanishing on the empty set. 
\end{theorem}

Hence, the core of any convex game is a deformed permutohedron, and each deformed permutohedron \(P\) is associated with the game \( (\calF_P, v_P) \) which is convex. 

\medskip 

A \emph{removahedron} is obtained by deleting inequalities in the facet description of the permutohedron. Then, each cooperahedron \( (\calF_P, v_P) \) with a coalition function satisfying the constraints of Equation~\eqref{eq: permutohedron-game} on \( \calF_p \) is a removahedron. 

\medskip 

The cooperahedron is also a generalization of the \emph{matroid polytope} \( P_M \) of a matroid \( M \), as shown by \textcite{ardila2010matroid}. Indeed, the matroid polytopes are a subclass of the deformed permutohedra. But what is interesting with the matroids is that they are characterized by a \emph{rank function} and a \emph{ground set}, the same way a game is characterized by a coalition function and a set of players. Hence, a matroid can be seen as a monotone convex game. Similar connections exist between convex games and the \emph{polymatroids} defined by \textcite{edmonds1970submodular}, as studied by \textcite{shapley1971cores}. 

\medskip 

Besides the connections made with polyhedral combinatorics, the cooperahedron is very useful in cooperative game theory, because numerous sets of solutions are cooperahedra, or properties of coalitions hold when a specific cooperahedron is nonempty. 

\section{Blind spots}\label{section: blind-spots}

This section aims to identify and characterize the preimputations that are dominated by a core element. The relation of domination between two preimputations can be restated using the concept of \emph{side payments}. 

\medskip 

A very particular feature of the games studied in this paper is that utility is transferable, i.e., can be sent from one player to another, and a unit of utility is worth the same for every player. These transfers are represented by side payments, i.e., \(n\)-dimensional vectors whose coordinates sum to zero. Like the preimputations, each coordinate of a side payment corresponds to a player and represents the change of its payment generated by the reallocation of utility the side payment induces. 

\medskip 

Geometrically, the set of side payments, denoted by \( \Sigma \) (or \( \Sigma_N \) if such clarification is needed), is the linear subspace of \(\bbR^N\) of codimension \(1\) parallel to the set of preimputations \(X(v)\), or equivalently the tangent space of \(X(v)\). 

\medskip 

Thus, the preimputation \(x\) dominates another preimputation \(y\) via a coalition \(S\) if there exists a side payment \( \sigma \in \Sigma \) such that 
\begin{equation} \label{eq: side-pay-dom}
\sigma(S) \leq e_S(y) \qquad \text{and} \qquad \sigma_i > 0, \forall \hspace{1pt} i \in S, 
\end{equation}
where \(e_S(y)\) is the \emph{excess} of \(S\) at \(y\) with respect to \( (N, v) \), given by \(e_S(y) \coloneqq v(S) - y(S)\). We write \( e_S^v(y) \) if the clarification is needed. 

\medskip 

The excess is the additional quantity of utility that a coalition can obtain by working on its own and refusing the offered preimputation. The balancedness property of balanced collections can be better understood using the following characterization. 

\begin{proposition}[\textcite{derks1998orderings}] \leavevmode \newline
A collection of coalitions \( \calC \) is balanced if and only if, for every side payment \( \sigma \in \Sigma \), 
\begin{itemize}
\item[\( \circ \)] either \( \sigma(S) = 0 \), for all \( S \in \calC \), 
\item[\( \circ \)] or there exists \( S, T \in \calC \) such that \( \sigma(S) > 0 \) and \( \sigma(T) < 0 \). 
\end{itemize}
\end{proposition}

In other words, it is impossible to increase the payment of a coalition in a balanced collection without decreasing the payment of another coalition, a property that resembles Pareto optimality. 

\begin{definition}
  A collection \( \calC \) of coalitions is \emph{unbalanced} if it does not contain a balanced collection. In particular, it is not itself a balanced collection. 
\end{definition}

It is important to notice that unbalancedness is not equivalent to the negation of balancedness. The collection \( \calC = \{ \{a, b\}, \{a, c\}, \{b, c\}, \{a\} \} \) is not balanced on \( \{a, b, c\} \), but \( \calC \setminus \{a\} = \{ \{a, b\}, \{a, c\}, \{b, c\} \} \) is. Collections that are neither unbalanced nor balanced are called \emph{weakly balanced} by \textcite{maschler1971kernel}. 

\medskip

The unbalanced collections were first studied in mathematical physics by \textcite{evans1992n}, while he was working on a specific hyperplane arrangement coming from quantum field theory. The same arrangement was coined \emph{resonance arrangement} by \textcite{shadrin2008chamber}, when they encountered it in algebraic geometry. Today, this arrangement is studied in various fields of mathematics. 

\medskip 

The deep connection between the unbalanced collections and the resonance arrangement comes from the fact that the connected components of the complement of the union of the hyperplanes are in bijection with the unbalanced collections. 

\medskip 
 
The characterization of the unbalanced collections using side payments demonstrates the relevance of these objects in cooperative game theory. 

\begin{proposition}[Billera, Moore, Moraites, Wang and Williams~\cite{billera2012maximal}]\label{prop: unbalanced} \leavevmode \newline 
A collection of coalitions \( \calC \) is unbalanced if and only if there exists a side payment \( \sigma \in \Sigma \) such that, for all \( S \in \calC \), we have \( \sigma(S) > 0 \). 
\end{proposition}

Unlike the balanced collections, the unbalanced collections are exactly the collections of coalitions the payments of which can all be improved at once. 

\medskip 

Now that we have identified for which collections there exists a side payment simultaneously increasing the payments of all the coalitions contained in them, we want to know for which unbalanced collections there exists such a side payment representing the domination via a coalition of the collection, as discussed in Equation~\eqref{eq: side-pay-dom}. 

\medskip 

We denote by \( X_\calC(v) \) the set of preimputations giving an unsatisfactory payment to all the coalitions in \( \calC \), and a satisfactory payment for all the other coalitions, i.e., 
\begin{equation*}
X_\calC(v) \coloneqq \{ x \in X(v) \mid x(S) < v(S) \text{ if and only if } S \in \calC \}. 
\end{equation*}

We say that the collection \( \calC \) is \emph{feasible} if its associated \emph{region} \( X_\calC(v) \) is nonempty. Note that the core is the region associated with the empty collection of coalition \( \calC = \emptyset \). Hence, the nonempty regions form a partition of the space of preimputations \( X(v) \). 

\begin{remark}
For any collection \( \calC \), the region \( X_\calC(v) \) is a cooperahedron.
\end{remark}

\begin{lemma}[\textcite{grabisch2021characterization}]\label{lemma: feasible} \leavevmode \newline 
  Let \( (N, v) \) be a balanced game. If \( \calC \) is a feasible collection, then it is unbalanced. 
\end{lemma}

Let \( \phi \) be the map associating to each preimputation \(x\), the set of coalitions 
\[
\phi(x) \coloneqq \{S \in \calN \mid x(S) < v(S)\}. 
\]
In other words, \( \phi(x) \) is the feasible collection associated with the region containing \(x\). 

\begin{definition}
  Let \(x\) be a preimputation and \(S\) be a coalition. We denote by \( \delta_S(x) \) the \emph{domination cone} of \(x\) with respect to \(S\), defined by 
  \[
  \delta_S(x) \coloneqq \{y \in X(v) \mid y_i > x_i, \hspace{2pt} \text{for all } i \in S\}. 
  \]
\end{definition}

The domination cone \( \delta_S(x) \) does not only include preimputations dominating \(x\) because some of them are not affordable for \(S\), as there exist some \( y \in \delta_S(x) \) such that \( y(S) > v(S) \). In particular, the set \( \delta_S(x) \) contains no preimputation dominating \(x\) whenever \(S\) does not belong to \( \phi(x) \). 

\begin{definition}
  Let \(x\) be a preimputation. We denote by \( \text{Aug}(x) \) the \emph{augmentation cone} of \(x\), defined by 
  \[
  \text{Aug}(x) \coloneqq \{y \in X(v) \mid y(S) \geq x(S), \hspace{2pt} \text{for all } S \in \phi(x)\}. 
  \]
\end{definition}

\begin{remark}
  For any preimputation \(x\) and any coalition \(S\), the domination cone \( \delta_S(x) \) and the augmentation cone \( \text{Aug}(x) \) are cooperahedra. 
\end{remark}

By Lemma~\ref{lemma: feasible}, a feasible collection is necessarily unbalanced, then by Proposition~\ref{prop: unbalanced}, the augmentation cone of any preimputation is nonempty. Let \( \zeta(x) \) be defined by 
\[
\zeta(x) \coloneqq \text{Aug}(x) \cap \left( \bigcup_{S \in \phi(x)} \delta_S(x) \right). 
\]

\begin{definition}
  Let \( \calC \) be a feasible collection. We say that \( \calC \) is \emph{inextricable} and that \( X_\calC(v) \) is a \emph{blind spot} if, for all \( x \in X_\calC(v) \), we have \( \zeta(x) = \emptyset \). 
\end{definition}

The importance of the sets \( \zeta(x) \) relies on the following result. 

\begin{lemma}
Let \( (N, v) \) be a balanced game. The core \( C(v) \) is a stable set only if, for all preimputations \(x\), we have \( \zeta(x) \neq \emptyset \). 
\end{lemma}

\begin{proof} 
  Assume that there exists \( x \in X(v) \) such that \( \zeta(x) = \emptyset \). Because the core is included in \( \text{Aug}(x) \), and \( \text{Aug}(x) \) does not intersect any set of preimputations dominating \(x\), the core cannot be externally stable and therefore is not stable. 
\end{proof}

\begin{proposition}
Let \( \calC \) be a feasible collection, and let \( y \in X_\calC(v) \) such that \( \zeta(y) \neq \emptyset \). Then, for all \( x \in X_\calC(v) \), we have \( \zeta(x) \neq \emptyset \). 
\end{proposition}

\begin{proof}
  Assume that there exists \( y \in X_\calC(v) \), \( z \in X(v) \) and a coalition \( S \in \calC \) such that \( z \in \text{Aug}(y) \cap \delta_S(y) \subseteq \zeta(y) \). Let \( x \in X_\calC(v) \). Denote by \( \sigma = x - y \) the side payment from \(y\) to \(x\). For all \(i \in S\), we have 
  \[
  {(z + \sigma)}_i = z_i + \sigma_i > y_i + \sigma_i = {(y + \sigma)}_i = x_i. 
  \]
  Hence, \( z + \sigma \in \delta_S(x) \). Moreover, for all \( T \in \calC \), we have 
  \[
  (z + \sigma)(T) = z(T) + \sigma(T) \geq y(T) + \sigma(T) = (y + \sigma)(T) = x(T),
  \]
  which implies that \( z + \sigma \in \text{Aug}(x) \cap \delta_S(x) \subseteq \zeta(x) \). 
\end{proof} 

The last result implies that the impossibility for a given state \(x\) to be renegotiated in another state \(y\) giving a better payment to all coalitions \( S \in \phi(x) \) while dominating \(x\) via a coalition in \(S\) only depends on the set \( \phi(x) \). In other words, the possibility to content a collection of coalitions with their payment does not depend on the values of the previous payments, but on the set system formed by the unsatisfied coalitions. 

\medskip 

However, the feasibility of a collection depends on \(v\), and can be checked using the non-emptiness characterization of cooperahedra. 

\medskip 

The next result is the most important of the section, because it characterizes the blind spots. 

\begin{theorem}\label{th: blind-spot}
  Let \( \calC \) be a feasible collection. Then \(X_\calC(v)\) is not a blind spot if and only if there exists a coalition \( S \in \calC \) such that \( \calC \cup \{ \{i\} \mid i \in S \} \) is unbalanced on \(N\).  
\end{theorem}

\begin{proof}
  Let \(x \in X_\calC(v)\), and denote by \(P\) the polyhedron \( \text{Aug}(x) \cap \delta_S(x) \). \(P\) being a cooperahedron, we denote by \( (\calF_P, v_P) \) the game associated with \(P\). Notice that \( \calF_P = \calC \cup \{ \{i\} \mid i \in S \} \) and, for all \( T \in \calF_P \), we have \( v_P(T) = x(T) \). 

  \medskip 

  We assume that \( P \neq \emptyset \). By Theorem~\ref{th: cooperahedron-non-empty}, \( (\calF_P, v_P) \) is balanced and the set \( \{ \{i\} \mid i \in S \} \) does not contain any effective coalition. Because \( \calC \) is feasible, by Lemma~\ref{lemma: feasible}, it is unbalanced. Therefore, if there exists a balanced collection \( \calB \subseteq \calF_P \), it must intersect \( \{ \{i\} \mid i \in S \} \). Moreover, 
  \[
  \sum_{T \in \calB} \lambda_S v_P(T) = \sum_{T \in \calB} \lambda_T x(T) = x(N) = v(N). 
  \]
  By Proposition~\ref{prop: effective}, \( \calB \subseteq \calE(\calF_P, v_P) \), contradicting Theorem~\ref{th: cooperahedron-non-empty}. Hence, \( \calF_P \) is unbalanced. 

  \medskip 

  Assume now that there is no balanced collection \( \calB \subseteq \calF_P \). Then, by Theorem~\ref{th: cooperahedron-non-empty}, the set \( P = \text{Aug}(x) \cap \delta_S(x) \) is nonempty, and for all \( x \in X_\calC(v) \), we have \( \zeta(x) \neq \emptyset \). 
\end{proof} 

\begin{example}
  Let us consider the flow game defined by the graph depicted in Figure~\ref{fig: flow-2}. The preimputation \(x = (2, 5, 5)\) is unsatisfactory for both coalition \( \{a, b\} \) and \( \{a, c\} \), but is satisfactory for all the other ones. Therefore, \( \calC = \{ \{a, b\}, \{a, c\} \} = \phi(y) \) is a feasible collection. Take \( S = \{a, b\} \). The set system \( \calF_S = \{ \{a, b\}, \{a, c\}, \{a\}, \{b\} \} \) contains the balanced collection \( \{ \{a, c\}, \{b\} \} \), therefore is not unbalanced. For \( T = \{a, c\} \), the set system \( \calF_T = \{ \{a, b\}, \{a, c\}, \{a\}, \{c\} \} \) contains \( \{ \{a, b\}, \{c\} \} \), and is not unbalanced. Then \( X_\calC(v) \) is a blind spot, and \( \calC = \{ \{a, b\}, \{a, c\} \} \) is inextricable. 
\end{example}

Theorem~\ref{th: blind-spot} gives a complete characterization of simultaneous improvements of the payments of a given collection of coalitions, by choosing a new state which dominates the previous one, while it only gives a partial characterization on stability. Indeed, if the initial state of the game lies in a blind spot, we should rationally expect the grand coalition not to form. The result is more general than just checking core stability, as it can be applied in a broader range of considerations. 

\section{Applications to market games}\label{section: market}

Let us illustrate the inextricability of collections of coalitions using \emph{market games}, defined by \textcite{shapley1969market}. First, any market game is balanced, so in this section we focus on the domination of an initial state by a core element.  

\medskip 

It is well-known that, in an exchange economy from the \textcite{arrow1954existence} model, the competitive equilibria always exist and belong to the core. However, it is still an open problem to characterize their stability, i.e., to know whether the economy will converge towards them. Even if there exist a lot of reasons to consider, we can observe that there exists some tension and dissatisfaction from some players and coalitions regarding the payment they received in real-life situations. 

\medskip 

Notwithstanding the similarities described above between the competitive equilibria and the core, the dynamic processes that aim to tend towards them are different. For the competitive equilibria, also known as Walrasian equilibria, the corresponding process consists of \emph{Walrasian auctions}, called the \emph{t{\^a}tonnement}. In our case, our \emph{cooperative equilibria} are sought by the domination process described earlier. 

\medskip 

In this regard, \textcite{shapley1955markets} said: ``Such `markets' are most commonly analyzed by the method of equilibrium points -- i.e., as if they were noncooperative games -- but there are some attractive reasons for using instead the `cooperative' theory of von Neumann and Morgenstern. For one thing, money transfers can be handled by side payments, without explicit determination of prices. Moreover, all bargaining can be regarded as part of the coalition-formation process, so that no formal rules for bids and offers, etc., are required''. This view agrees with the critics of \textcite{maskin2016can} formulated more than 60 years after, about the robustness of cooperative games and the relevance of the study of coalition formation, that we discussed in the introduction.

\medskip 

Moreover, in the case of non-convex preferences of the players, the core of the economy may be nonempty even if there are no competitive equilibria, as shown by \textcite{shapley1963core} and the references therein. Hence, the study of the core and the domination relation is relevant to a broader class of economic situations. 

\medskip

To define market games and start the investigation of cooperative equilibria, we use the model of mathematical markets described by \textcite{shapley1969market}. 

\begin{definition}
A \emph{market} is denoted by the symbol \((N, G, A, U)\) where
\begin{itemize}
  \item[\( \circ \)] \(N\) is a finite set of \emph{players}, 
  \item[\( \circ \)] \(G\) is the nonnegative orthant of a finite-dimensional vector space, called the \emph{commodity space},
  \item[\( \circ \)] \( A = \{ a^i \mid i \in N \} \) is an indexed collection of points in \(G\), called the \emph{initial endowments}, 
  \item[\( \circ \)] \( U = \{ u^i \mid i \in N \} \) is an indexed collection of continuous, concave functions from \(G\) to the real numbers, called the \emph{utility functions}. 
\end{itemize}
\end{definition}

In this market, the players of \(N\) aim to exchange some of their initial endowments for commodities they prefer. Their preferences are represented by the utility functions, which evaluate each bundle of commodities with real numbers, which are totally ordered and allow for comparisons of any bundles.

\medskip 

In a certain coalition \(S\), if cooperation occurs, the players put all their initial endowments, and seek for a reallocation of the commodities \( x = \{x^i \mid i \in S\} \subseteq G^S \). Because cooperation is restrained to the coalition \(S\), the reallocation have to satisfy the condition \( \sum_{i \in S} x^i = \sum_{i \in S} a^i \). We denote by \( X^S \) the set of these reallocations. Without considering any help from outside the coalition, the players of \(S\) can acquire an amount of utility given by 
\[
v(S) = \max_{x \in X^S} \sum_{i \in S} u^i(x). 
\] 
The collection of the numbers \( \{ v(S) \mid S \in \calN \} \) defines the \emph{market game} \((N, v)\) associated with the market \((N, G, A, U)\). 

\medskip 

The study of market games is without any loss of generality for the study of balanced games. Indeed, \textcite{shapley1969market} have demonstrated that any balanced game can be transformed into a market game while preserving all the domination relation between any pair of preimputations. Using market games allows for simpler description of games, requiring less information than a naive enumeration of \(2^n - 1\) real numbers, and gives a more intuitive view on games, as we have previously done with flow games. 

\medskip 

Together with a market game \((N, v)\), consider an initial state \(x\). The \(n\)-dimensional vector \(x'\) defined, for all \(i \in N\), by \( x'_i = u^i(a^i) \) is not necessarily a preimputation, but satisfies \( x'(N) \leq v(N) \) by definition of a market game. Hence, the projection \(x\) of \(x'\) onto the space of preimputations \(X(v)\), given by 
\[
  x_i = x'_i + \frac{v(N) - x'(N)}{n},
\]
is a valid initial state, which clearly dominates \(x'\) via any possible coalition when \(x' \neq x\). We consider \(x\) to be outside the core for the present study. The question is whether the grand coalition will form, or, equivalently, if there exists some coalitions which would prefer to defect. Given the initial state, it is unclear whether there exists a core element dominating it with only the data provided by \(A\) and \(U\). 

\medskip 

The relevant information for this investigation is the feasible collection \(\phi(x)\) determined by the initial state. By applying the Theorem~\ref{th: blind-spot} and using the algorithms developed by the author, Grabisch and Sudh{\"o}lter~\cite{laplace2023minimal}, it is possible to know whether \(x\) is in a blind spot, and whether \(\phi(x)\) is inextricable. 

\medskip 

The commodities defining the set \(G\) need not be physical goods. They can be the amount of work, expressed in units of time for instance, that a player is ready to offer to the market. In this case, the player always has the possibility to use the threat of domination at any time, because she can easily withdraw her initial endowments. This threat being physically feasible, by Theorem~\ref{th: force-pay}, any coalition can demand a dominating state and can rationally expect to get satisfaction, provided that there does not exist another coalition demanding a contradictory state. 

\medskip 

\textcite[p.~150]{shubik1982game} stated that ``a game that has a core has less potential for social conflict than one without a core''. For similar reasons, we can probably state that, when the current state of a game, a market, or an economy, is not dominated by a core element, the potential of social conflict increases. 

\medskip 

Moreover, von Neumann and Morgenstern proposed that a stable set is viewed as ``a \emph{standard of behavior} -- or a \emph{tradition}, \emph{social convention}, \emph{canon of orthodoxy}, or \emph{ethical norm} -- against which any contemplated outcome can be tested'' \cite[p.~161]{shubik1982game}. Indeed, an allocation outside the stable set is dominated by an allocation inside it, and then, because of Theorem~\ref{th: force-pay}, the state of the game evolve towards the stable set. 

\medskip 

Studying the coincidence between the core, which contains the allocations satisfying a fairness property, and the stable sets, representing the social conventions of the society under study, reveals how this aforementioned society unfairly treats the coalitions belonging to an inextricable collection for a given current state. 


\section{Concluding remarks}

In this paper, we have introduced a novel approach to consider coalition formation. Usually, the grand coalition is assumed to form, or, in the best case, to form whenever the core is nonempty. We have seen with several examples that these assumptions cannot hold. Instead, we considered that the grand coalition forms if the initial state of the game is either in the core, or is core dominated. The games for which the grand coalition forms for any possible initial state are thus the games with a stable core. When the initial state is not included in the core, it usually aggrieves several coalitions, with possibly divergent interests. 

\medskip 

During the study of the feasibility of a simultaneous improvement of the payments of these coalitions, we identify a specific polyhedron which frequently appears. We call these polyhedra \emph{cooperahedra}. We give a characterization of their nonemptiness, by generalizing the Bondareva-Shapley Theorem. 

\medskip 

Then, the main theorem of the paper formalizes an equivalence between the intertwinement of the considered coalitions and the possibility to simultaneously improve their payments. This result identifies blind spots, which are sets of initial states that cannot be core dominated. This gives a better understanding of the convergence of economic or social environments towards an outcome benefitting everyone. 

\printbibliography{}

\end{document}